\documentclass[twocolumn,prl]{revtex4}
\usepackage{graphicx,amsmath}

\begin{document}
\title{Cluster States for Continuous-Variable Multipartite Entanglement}
\author{Jing Zhang}
\affiliation{State Key Laboratory of Quantum Optics and\\
Quantum Optics Devices, Institute of Opto-Electronics, Shanxi University,\\
Taiyuan 030006, P.R.China\\
E-mail: jzhang74@yahoo.com, jzhang74@sxu.edu.cn}

\begin{abstract}
We introduce a new class of continuous-variable (CV) multipartite
entangled states, the CV cluster states, which might be generated
from squeezing and kerr-like interaction $H_I=
\rlap{\protect\rule[1.1ex]{.325em}{.1ex}}h%
\chi X_AX_B$. The entanglement properties of these states are
studied in terms of classical communication and local operations.
The quantum teleportation network with cluster states is
investigated. The graph states as the general forms of cluster
states are presented, which may be used to generate CV
Greenberger-Horne-Zeilinger states by simply local measurements and
classical communication. A chain for one-dimensional example of
cluster states can be readily experimentally produced only with
squeezed light and beamsplitters.
\end{abstract}


\maketitle

Entanglement is the most fascinating features of quantum mechanics
and plays a central role in quantum information processing. In
recent years, there has been an ongoing effort to characterize
qualitatively and quantitatively the entanglement properties of
multiparticle systems and apply then in quantum communication and
information. The study of multipartite entangled states in discrete
variable regime have shown that there exist different types of
entanglement, inequivalent up to local operations and classical communication%
\cite{one} (LOCC). For example, it is now well known that the
Greenberger-Horne-Zeilinger (GHZ) and W states are not equivalent up
to LOCC. Recently, Briegel and Raussendorf introduced a special kind
of multipartite entangled states, the so-called cluster states,
which can be created via an Ising Hamitonian\cite{two}. It has been
shown that via cluster states, one can implement a quantum computer
on a lattice of qubits. In this proposal, which is known as ``one
way quantum computer'', information is written onto the cluster and
then is processed and read out from the cluster by one-qubit
measurements\cite{three}.

In recent years, quantum communication, or more general quantum
information with continuous variables (CV) has attracted a lot of
interest and appears to yield very promising perspectives concerning
both experimental realizations and general theoretical
insights\cite{three1}, due to relative simplicity and high
efficiency in the generation, manipulation, and detection of CV
state. The investigation of CV multipartite entangled states has
made significant progress in theory and experiment. The
quantification and scaling of multipartite entanglement has been
developed by the different methods, for example, by determining the
necessary and sufficient criteria for their
separability\cite{four,five,five1}, by the entanglement of
formation\cite{six}. Multipartite quantum protocols were proposed
and performed experimentally
such as quantum teleportation network\cite{seven}, controlled dense coding%
\cite{eight} and quantum secret sharing\cite{nine} based on
tripartite entanglement. The study of entanglement properties of the
harmonic chain has aroused great interest\cite{ten}, which is in
direct analogy to spin chain with an Ising interaction. Furthermore
CV entanglement swapping has been demonstrated
experimently\cite{twelve}, in which a four-partite entangled state
may be generated. To date, the study and application of CV
multipartite entangled states mainly focused on CV GHZ-type states.
However,the richer understanding and more definite classification to
CV multipartite entangled states are needed for developing CV
quantum information network. In this letter, we introduce a new
class of CV multipartite entangled states - the CV cluster-type
states, which is different from CV GHZ-type states. We propose an
interaction model to realize CV cluster-type states and compare
these states to GHZ-type states in terms of LOCC.

We consider $N$-mode $N$-party described by a set of quadrature operators $%
R=(X_1,Y_1;X_2,Y_2,...;X_1,Y_n)$ obeying the canonical commutation relation $%
[X_l,Y_k]=i\delta _{lk}$. The Kerr-like interaction Hamiltonian is employed
with the form $H_I=
\rlap{\protect\rule[1.1ex]{.325em}{.1ex}}h%
\chi X_lX_k$, so in this process quadrature-amplitude (``position'')
and quadrature-phase (``momentum'') operators are transformed in
Heisenberg picture according to the following expressions:
\begin{eqnarray}
X_l^{\prime } &=&X_l,\quad Y_l^{\prime }=Y_l+gX_k,  \label{Heis} \\
X_k^{\prime } &=&X_k,\quad Y_k^{\prime }=Y_k+gX_l,  \nonumber
\end{eqnarray}
where $g=-\chi t$ is the gain of the interaction, $\chi$ and $t$ are
the Kerr nonlinear coefficient and the interaction time
respectively. The important feature of this Hamiltonian is that the
momentum $Y_l$ and $Y_k$ pick up the information of the position
$X_k$ and $X_l$ respectively, while position keep unchanging. The
Kerr-like interaction between light was widely investigated in many
previous experiments\cite{thirteen} and recently, has been
experimentally observed between the light and collective spin of
atomic samples\cite{forteen,forteen1}.

Considering first the one-dimensional example of a chain of $N$
modes which are numbered from $1$ to $N$ with next-neighbor
interaction. Initially, all
modes are prepared in the quadrature-phase squeezed state $%
X_l=e^{+r}X_l^{(0)},Y_l=e^{-r}Y_l^{(0)}$, where $r$ is the squeezing
parameter and a superscript $``(0)"$ denotes initial vacuum modes. Applying
the Kerr-like interactions to next-neighbor modes of a chain at the
different time or the same time yields a CV cluster-type state in the form
\begin{eqnarray}
X_1^C &=&e^{+r}X_1^{(0)},\quad Y_1^C=e^{-r}Y_1^{(0)}+e^{+r}X_2^{(0)},
\label{cluster} \\
X_2^C &=&e^{+r}X_2^{(0)},\quad
Y_2^C=e^{-r}Y_2^{(0)}+e^{+r}X_1^{(0)}+e^{+r}X_3^{(0)},  \nonumber \\
&&\cdots   \nonumber \\
X_i^C &=&e^{+r}X_i^{(0)},\quad
Y_i^C=e^{-r}Y_i^{(0)}+e^{+r}X_{i-1}^{(0)}+e^{+r}X_{i+1}^{(0)},  \nonumber \\
&&\cdots   \nonumber \\
X_{N-1}^C &=&e^{+r}X_{N-1}^{(0)},\quad
Y_i^C=e^{-r}Y_{N-1}^{(0)}+e^{+r}X_{N-2}^{(0)}+e^{+r}X_N^{(0)},  \nonumber \\
X_N^C &=&e^{+r}X_N^{(0)},\quad Y_i^C=e^{-r}Y_N^{(0)}+e^{+r}X_{N-1}^{(0)}.
\nonumber
\end{eqnarray}
Here, without loss of generality, we put the gain of the interaction $%
g=1$. For simplicity we discuss the properties of CV cluster-type
states in the ideal case $r\rightarrow \infty $ corresponding to
infinite squeezing.
When $N=2$, one obtains the state with total position $X_1^C+X_2^C%
\rightarrow 0$ and relative momentum $Y_1^C-Y_2^C\rightarrow 0$ by
applying a local $-90^0$ rotation transformation $X^C\rightarrow
Y^C$, $Y^C\rightarrow -X^C$ on mode 2, which corresponds to
Einstein-Podolsky-Rosen (EPR) or two-mode squeezed states. The
entanglement of a state is not affected by local unitary
transformation. It obviously exhibits maximum bipartite
entanglement. Similarly, one obtains the state for $N=3$ with total
position
$X_1^C+X_2^C+X_3^C\rightarrow 0$ and relative momentum $Y_i^C-Y_j^C%
\rightarrow 0$ $(i,j=1,2,3)$, which corresponds to CV three-mode
GHZ-like states\cite{seven}. However, quantum correlation of
position and momentum of
the state for $N=4$ case become $X_1^C+X_2^C+X_3^C\rightarrow 0$, $%
X_3^C+X_4^C\rightarrow 0$and $Y_1^C-Y_2^C\rightarrow 0$, $%
Y_2^C-Y_3^C+Y_4^C\rightarrow 0$ by applying a local $-90^0$ rotation
transformation on modes $2$ and $4$. Apparently, CV four-partite
cluster-type state is not equivalent to a GHZ-like state with total
position
$X_1^C+X_2^C+X_3^C+X_4^C\rightarrow 0$ and relative momentum $%
Y_i^C-Y_j^C\rightarrow 0$ $(i,j=1,2,3,4)$. More generally, CV
$N$-partite cluster-type states and GHZ-type states are not
equivalent for $N>3$, as we shall see below.

We now compare the entanglement properties of CV cluster and
GHZ-type states by LOCC in the limit of infinite squeezing. First,
we discuss the persistency of entanglement of an entangled state of
$N$-partite that means the minimum number of local measurements such
that, for all measurement outcomes, the state is completely
disentangled\cite{two}. An explicit strategy to disentangle CV
cluster-type state (2) is to measure position of all even number
parties, $j=2,4,6,\cdots $, then displace momentum of the remaining
(unmeasured) parties with the measured results, such as momentum of
partite $1$ displaced as $Y_1^{\prime C}=Y_1^C-X_2^C$
$=e^{-r}Y_1^{(0)}$ by measured result $X_2^C$ of partite $2$;
partite $3$ as $Y_3^{\prime C}=Y_3^C-X_2^C-X_4^C$ by $2,4$. It is
obvious that the remaining parties become originally prepared
quadrature-phase squeezed state that are a product state and
completely unentangled. Thus the minimum number to disentangle
cluster state is integer part$[N/2]$. For the GHZ state, in
contrast, a single local measurement suffices to bring it into a
product state. From this point of view, it is impossible to destroy
all the entanglement of the cluster if less than integer part$[N/2]$
parties are traced out (discarded). But only if one partite for GHZ
state is traced, the remaining state will be completely unentangled.
Note that it is not true in the case of finite squeezing. For
example, if three weakly squeezed vacuum states are used to generate
tripartite GHZ state, the state is a fully inseparable tripartite
entangled state, but the remaining bipartite state after tracing out
one of the three subsystems is still entangled\cite {seven,eight}.

Next, we investigate the entanglement property by the quantum teleportation
network. In other words, we will answer the question that how many parties
to be measured from $N$-partite entangled state may make bipartite
entanglement between any two of $N$ parties ``distilled'', which enables
quantum teleportation. We first show that the parties at the ends of chain,
i.e., party $1$ and $N$, can be brought into a bipartite entanglement by
measuring the parties $2$, $\ldots $, $N-1$. After measurement of momentum
of party $2$, the remaining parties are identical to an entangled chain of
length $N-1$ when one displaces the position of party $1$ as $X_1^{\prime
C}=X_1^C-Y_2^C$ by measured result $Y_2^C$ of partite $2$, then applies a
local rotation transformation $X^C\rightarrow -Y^C$, $Y^C\rightarrow X^C$ on
party $1$. We can repeat this procedure and measure party $3$, and so on. At
the end, party $1$ and $N$ are brought into a bipartite entanglement. To
bring any two party $j$, $k$ $(j<k)$ from the chain $\{1,2,\ldots ,N\}$ into
a bipartite entanglement, we first measure position of the ``outer'' parties
$j-1$ and $k+1$, then displace momentum of parties $j$ and $k$ as $%
Y_j^{\prime C}=Y_j^C-X_{j-1}^C$ and $Y_k^{\prime C}=Y_k^C-X_{k+1}^C$, which
projects the parties $j,j+1,\ldots ,k$ into an entangled chain of length $%
K-j+1$. This process is called ``disconnection'' that break a chain into
parts, for example, one breaks a chain into two independent chains when
measuring position of party $j$ and displacing momentum of parties $j-1$ and
$j+1$ as $Y_{j-1}^{\prime C}=Y_{j-1}^C-X_j^C$ and $Y_{j+1}^{\prime
C}=Y_{j+1}^C-X_j^C$. A subsequent measurement of the ``inner'' parties $%
j+1,\ldots ,k-1$ will then project parties $j$, $k$ into a bipartite
entanglement, as shown previously. Note that the assistance of all the
``inner'' parties is necessary to bring any two party from the chain into a
bipartite entanglement, however, for the ``outer'' parties it depends on
which strategy is chosen. The optimum strategy as shown previously is to
measure next-neighbor parties $j-1$ and $k+1$. The other strategies may be
with the help of parties $j-2,j-3$ and $k+2,k+3$, or $j-2,j-4,j-5$ and $%
k+2,k+4,k+5,$ et.al. For the $N$-partite GHZ state, in contrast, bipartite
entanglement between any two of the $N$ parties is generated with the help
of local position measurement of $N-2$ modes\cite{seven}.

In the following we will generalize one-dimensional CV cluster to
graph states that correspond to mathematical graphs, where the
vertices of the graph play the role of quantum physical systems,
i.e. the individual modes and the edges represent interactions. A
graph $G=(V,E)$ is a pair of a finite set of $n$ vertices $V$ and a
set of edges $E$, the elements of which are subsets of $V$ with two
element each\cite{fifteen}. We will only consider simple graphs,
which are graphs, that contain neither loops (edges
connecting vertices with itself) nor multiple edges. When the vertices $%
a,b\in V$ are the endpoints of an edge, they are referred to as being
adjacent. An $\left\{ a,c\right\} $-path is a order list of vertices $%
a=a_1,a_2,\ldots ,a_{n-1},a_n=c$, such that for all $i$, $a_i$ and $a_{i+1}$
are adjacent. A connected graph is a graph that has an $\left\{ a,c\right\} $%
-path for any two $a,c\in V$. Otherwise it is referred to as
disconnected. The neighborhood $N_a\subset V$ is defined as the set
of vertices $b$ for which $\left\{ a,b\right\} \in E$. In other
words, the neighborhood is the set of vertices adjacent to a given
vertex. The CV\ graph states with a given graph $G$ are conveniently
defined as
\begin{eqnarray}
X_a^G &=&e^{+r}X_a^{(0)},\quad
Y_a^G=e^{-r}Y_a^{(0)}+e^{+r}\sum_{b\in N_a}X_b^{(0)},\qquad
\label{graph} \\
\text{for }a \in V. \nonumber
\end{eqnarray}
Equation (\ref{graph}) can be used to generalize some of the
entanglement properties from the 1D case to graph states. To bring
any two parties on sites $a,b\in V$ into a bipartite entanglement,
we first select a one-dimensional path $P\subset V$ that connects
sites $a$ and $b$. Then we measure all neighboring modes surrounding
this path in the position component. By this procedure, we project
the parties on the path $P$ into a state, that is up to local
displacement on the momentum with measured results, identical to the
linear chain Eq.(\ref{cluster}). We have thereby reduced the graph
states to the one-dimensional problem.

%
%
\begin{figure}
\centerline{
\includegraphics[width=2.6in]{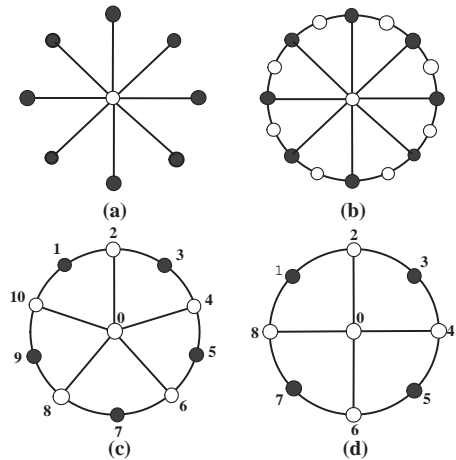}
} \vspace{0.1in}
\caption{ Generation of a CV GHZ state from graph state in the form
of ring+star. The solid parties are projected into a CV GHZ state by
measuring the circled parties in the appropriate basis and by
subsequent local displacement. \label{Fig_ring} }
\end{figure}

Entanglement is regarded as a resource for quantum information and
thus the question arises which states can be obtained from graph
states by local operations and classical communication. Here, we
find the CV GHZ states of the subset of parties $V^{\prime }\subset
V$ can be obtained from graph states of the special form: star and
ring+star by LOCC as shown in Fig.1. For the simple case of star
structure as in Fig.1a, we only need to measure momentum of the
central party, then displace position of only one of remaining
parties as $X_1^{\prime G}=X_1^G-Y_0^G$. Apparently, CV GHZ state is
generated with total position $X_1^{\prime G}+X_2^{\prime G}+\ldots
+X_8^{\prime G}\rightarrow 0$ and relative momentum $Y_i^{\prime
G}-Y_j^{\prime G}\rightarrow 0$ $(i,j=1,2,\ldots ,8)$. The case of
star+ring structure as in Fig.1b may be reduced to the case of
Fig.1a by disconnection operation on the circled parties in the
ring. However, the cases for Fig.1c and d are completely distinct
from the previous cases. For the case in Fig.1c, after measuring
momentum of the circled partes, we displace momentum of only
one of remaining parties with measured result of central party as $%
Y_1^{\prime G}=Y_1^G-2Y_0^G$ and position of remaining odd parties with
measured momentum of even parties in the ring as $X_1^{\prime G}=X_1^G$, $%
X_3^{\prime
G}=X_3^G-Y_4^G+Y_6^G-Y_8^G+Y_{10}^G=X_1^G-e^{-r}(Y_4^{(0)}-Y_6^{(0)}+Y_8^{(0)}-Y_{10}^{(0)})
$, $X_5^{\prime G}=X_5^G+Y_2^G-Y_4^G$, $X_7^{\prime G}=X_7^G-Y_8^G+Y_{10}^G$%
, $X_9^{\prime G}=X_9^G+Y_2^G-Y_4^G+Y_6^G-Y_8^G$. Thus we obtain CV GHZ
state with total momentum $Y_1^{\prime G}+Y_3^{\prime G}+Y_5^{\prime
G}+Y_7^{\prime G}+Y_9^{\prime G}\rightarrow 0$ and relative position $%
X_i^{\prime G}-X_j^{\prime G}\rightarrow 0$ $(i,j=1,3,\ldots ,9)$. But CV
GHZ state can not be generated from the case in Fig.1d. In fact, $N_{ring}^m$
measured results of circled parties in the ring correspond to $N_{ring}^m$
linear algebraic equations. One of $N_{ring}^m$ equations is a linear
combination of the others (which is called degeneracy) when $N_{ring}^m$ is
even, and all equations are linearly independent when odd. Hence, the case
for the odd number of measured parties in the ring can be projected to GHZ
state and the case of the even number can not.

We briefly mention that, to embody Eq.(\ref{cluster}) and (\ref{graph}), we
can use the off-resonant interaction of linearly polarized optical buses
with $N$ ensembles of atoms (confined in vapor cell), providing the
Hamiltonian $H_{OA}=
\rlap{\protect\rule[1.1ex]{.325em}{.1ex}}h%
\chi X_OX_A$. Here, $X_O$ $(X_A)$ is the position operator of the
bus (atomic ensemble). First, we prepare the spin-squeezed atomic
ensembles by means of Kerr-like interaction $H_{OA}$ and projection
measurement on light\cite
{forteen,sixteen}. In the second step, the Kerr-like interaction $%
H_{A_{lk}}= 
\rlap{\protect\rule[1.1ex]{.325em}{.1ex}}h%
\chi^{\prime }X_{A_l}X_{A_k}$between ensembles can be implemented by
first letting a optical bus interact with ensemble $l$, then
applying $90^0$ rotation transformation on the optical bus, then
letting the optical bus interact with ensemble $k$, then applying
$-90^0$ rotation transformation on
the optical bus, and finally letting the optical bus interact with ensemble $%
l$.

%
%
\begin{figure}
\centerline{
\includegraphics[width=3.3in]{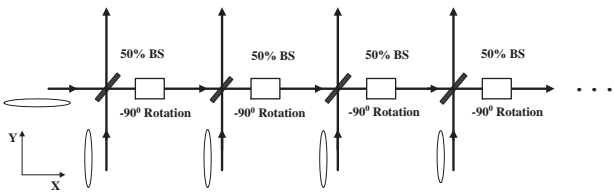}
} \vspace{0.1in}
\caption{ A schematic diagram of generating a chain by means of
squeezed light and beamsplitters. \label{Fig_BS} }
\end{figure}

Stimulating opportunities also come from the experimental
demonstration of CV entanglement swapping demonstrated\cite{twelve},
in which a four-partite entangled state may be generated. Here we
show a chain for one-dimensional example of cluster states can be
readily experimentally produced only with squeezed light and
beamsplitters as shown in Fig.2. A chain emerges from a sequence of
$N-1$ 50\% beam splitters and $-90^0$ rotations with $N$ inputs of
squeezed states. By choosing the squeezing direction of one distinct
input mode orthogonal to that of the remaining input modes (mode 1
squeezed in momentum and the other modes squeezed in position as
shown in Fig.2). For example, the quantum correlation of position
and momentum of CV chain states for $N=4$
case become $\sqrt{2}X_1^C+X_2^C+\sqrt{2}X_3^C\rightarrow 0$, $%
X_3^C+X_4^C\rightarrow 0$ and $Y_1^C-\sqrt{2}Y_2^C\rightarrow 0$, $\sqrt{2}%
Y_2^C-Y_3^C+Y_4^C\rightarrow 0$ by applying a local $-90^0$ rotation
transformation on modes $2$ and $4$. We see the quantum correlation
of CV chain states generated by squeezed light and beamsplitters is
equivalent to that of Eq.(\ref{cluster}), although the factors in
quantum correlation are different. Note that the beamsplitter
operation only is used to connect vertices one by one for generating
one-dimensional cluster states, which can not be used to connect two
chains together or generate graph states.

In conclusion, we have introduced a new class of CV multipartite
entangled states. We have proposed an interaction model to realize
CV cluster-type states and compared these states to GHZ-type states
in terms of LOCC. Those states are different with GHZ states and the
entanglement of the states is harder to be to destroy than GHZ
states. The set consisting of measured parties for distilling
bipartite entanglement from any pair of $N$ parties depends on where
they lie in the cluster. We have generalized one-dimensional CV
cluster to graph states. The graph states can be regarded as a
resource for other multipartite entangled states, such as GHZ
states. To embody CV cluster states, we have described the
experimental setups of atomic ensembles and squeezed light that
offer nice perspectives in the study of CV multipartite entangled
states. Cluster states for CV multipartite entanglement may be
applied to multipartite quantum protocols and are of practical
importance in realizing more complicated quantum computation and
quantum communication among many parties.

{\bf ACKNOWLEDGMENTS}

J. Zhang thanks K. Peng, C. Xie, T. Zhang, J. Gao, and Q. Pan for
the helpful discussions. This research was supported in part by
National Natural Science Foundation of China (Approval No.60178012),
Program for New Century Excellent Talents in University, and Shaxi
Province Young Science Foundation (Approval No.20021014).


\begin{references}
\bibitem{one}  W. D$\ddot{u}$r, G. Vidal, and J. I. Cirac, Phys. Rev. A \textbf{62},
062314 (2002); F. Verstraete, J. Dehaene, B. De Moor, H. Verschelde,
Phys. Rev. A \textbf{65}, 052112 (2002).

\bibitem{two}  H. -J. Briegel and R. Raussendorf, Phys. Rev. Lett. \textbf{86}, 910
(2001).

\bibitem{three}  R. Raussendorf and H. -J. Briegel, Phys. Rev. Lett. \textbf{86}, 5188
(2001).

\bibitem{three1}  S. L. Braunstein and A. K. Pati, \emph{Quantum Information with Continuous
Variables} (Kluwer Academic, Dordrecht, 2003).

\bibitem{four}  G. Giedke, B. Kraus, M. Lewenstein, and J. I. Cirac, Phys.
Rev. A \textbf{64}, 052303 (2001).

\bibitem{five}  P. van Loock, and A. Furusawa, Phys. Rev. A \textbf{67}, 052315
(2003).

\bibitem{five1}  G. Adesso, A. Serafini, and F. Illuminati, Phys. Rev. Lett.
\textbf{93}, 220504 (2004).

\bibitem{six}  M. M. Wolf, F. Verstraete, and J. I. Cirac, Phys. Rev. Lett.
\textbf{92}, 087903 (2004).

\bibitem{seven}  P. van Loock and S. L. Braunstein, Phys. Rev. Lett. 84,
3482 (2000); H. Yonezawa, T. Aoki, A. Furusawa, Nature \textbf{431},
430 (2004).

\bibitem{eight}  J. Zhang, C. Xie, K. Peng, Phys. Rev. A {\bf 66},
032318 (2002); J. Jing, J. Zhang, Y. Yan, F. Zhao, C. Xie, K. Peng,
Phys. Rev. Lett. {\bf 90}, 167903 (2003).

\bibitem{nine}  A. M. Lance, T. Symul, W. P. Bowen, B. C. Sanders and P. K.
Lam, Phys. Rev. Lett. \textbf{92}, 177903 (2004).

\bibitem{ten}  K. Audenaert, J. Eisert, M. B. Plenio, Phys. Rev. A \textbf{66},
042327 (2002); A. Botero and B. Reznik, Phys. Rev. A \textbf{70},
052329 (2004).

\bibitem{twelve}  J. Zhang, C. Xie, and K. Peng, Phys. Lett. A \textbf{299}, 427
(2002); X. Jia, X. Su, Q. Pan, J. Gao, C. Xie, K. Peng, Phys. Rev.
Lett. \textbf{93}, 250503 (2004).

\bibitem{thirteen}  S. F. Pereira et al., Phys. Rev. Lett. \textbf{72}, 214 (1994);
K. Bencheikh et al., ibid \textbf{75}, 3422 (1995); R. Bruckmeier et
al., ibid. \textbf{78}, 1243 (1997); P. Grangier et al., Nature
(London) \textbf{396}, 537 (1998).

\bibitem{forteen}  L. -M. Duan et al., Phys. Rev. Lett. \textbf{85}, 5643 (2000); A.\
Kuzmich and E. S. Polzik, ibid. \textbf{85}, 5639 (2000); B.
Julsgaard et al., Nature (London) \textbf{413}, 400 (2001); C.
Schori et al., Phys. Rev. Lett. \textbf{89}, 057903 (2002).

\bibitem{forteen1}  J. Zhang, K. Peng, and S. L. Braunstein, Phys. Rev. A \textbf{68},
035802 (2003).

\bibitem{fifteen}  M. Hein, J. Eisert, and H. J. Briegel, Phys. Rev. A \textbf{69},
062311 (2003).

\bibitem{sixteen}  J. M. Geremia, J. Stockton, H. Mabuchi, Science \textbf{304}, 270
(2004).
\end{references}
\end{document}